\documentclass[12pt]{article}

\usepackage{soul}

\usepackage{tikz}
\usepackage{epsfig,latexsym,amsfonts,amsmath,amsthm,amssymb,amsbsy,multirow,slashed,wasysym,textcomp,subfigure,wrapfig,datetime,comment,mathtools,cancel,cite,twistor,mathrsfs}
\usepackage[hidelinks]{hyperref}
\usepackage[font={footnotesize},bf]{caption}

\newcommand{\csch}{\mathrm{csch}}

\usepackage[normalem]{ulem} 
\numberwithin{equation}{section}
\def\ee{\end{equation}}
\def\be{\begin{equation}}
\def\bea{\begin{eqnarray}}
\def\eea{\end{eqnarray}}
\newcommand{\beq}{\begin{eqnarray}}
\newcommand{\eqq}{\end{eqnarray}}
 \newcommand{\badat}{\begin{alignedat}}
 \newcommand{\eadat}{\end{alignedat}}

\newcommand{\eal}[1]{\be \begin{aligned} #1 \end{aligned}\end{equation}} 
\newcommand{\sech}{\mathrm{sech}}
\newcommand{\eqn}[1]{\be #1 \end{equation}} 
\newcommand{\eqa}[1]{\bea  #1\end{eqnarray}}

\renewcommand{\d}{\mathrm{d}}

\long\def\new#1\endnew{{\bf #1}}		
\long\def\del#1\enddel{}
\def\eps{\epsilon }
\def\del{\partial}

\def\re{\mathrm{e}}

\def\s{\sigma }
\def\bs{\bar{\sigma}}

\def\l{\lambda }

\usepackage{color}

\newcommand{\pink}[1]{\textcolor{\pink}{#1}}

\definecolor{dblue}{rgb}{0.2,0.50,0.80}

\newcommand{\bPhi}{{\bf\Phi}}
\usepackage{xspace}

\def\S{\mathcal{S}}

\def\O{\mathcal{O}}

\def\D{{\Delta}}

\def\G{{\Gamma}}

\def\sllr{$\text{SL}(2,\mathbb{R})\times \overline{\text{SL}}(2,\mathbb{R})$}
\newcommand{\sot}{$\mathrm{SO}(2,2)$\xspace}

\def\AdS{{${\rm AdS}_3/{\mathbb{Z}}$}}

\newcommand{\ket}[1]{\left|#1\right\rangle}
\newcommand{\bra}[1]{\left\langle#1\right|}

\def\bs{ {\bar\sigma} }
\def\s{ {\sigma} }

\def\t2{T$^{1,1}$}
\def\d{\delta}

 \def\e{\epsilon}

\def\hs{{\hat \s}}
\def\hbs{{\hat \bs}}
\catcode`,\active
\newcommand*\pFq{\begingroup
        \catcode`\,\active
        \def ,{\mskip\pFqskip\relax}%
        \dopFq
}

\catcode`\,12
\def\dopFq#1#2#3#4#5{%
        {}_{#1}F_{#2}\biggl[\left.\genfrac..{0pt}{}{#3}{#4}\right|#5\biggr]%
        \endgroup
}

\begin{document}
\begin{titlepage}
\unitlength = 1mm~\\
\vskip 3cm
\begin{center}

{\LARGE{ Quantum Fields on Time-Periodic AdS$_3/\mathbb{Z}$}}

\vspace{0.8cm}
Walker Melton$^{*\dagger}$, Andrew Strominger$^*$ and Tianli Wang$^*$\\
\vspace{1cm}

{\it  $^*$Center for the Fundamental Laws of Nature, Harvard University, Cambridge, MA 02138, USA\newline
$^\dagger$Society of Fellows, Harvard University, Cambridge, MA 02138, USA} \\

\vspace{0.8cm}

\begin{abstract}
We consider  a  free complex massive scalar on the quotient spacetime \AdS, which has the  isometry group \sot  rather than its  universal cover. This problem is of  interest as a special example of QFT on a spacetime with closed timelike curves (CTCs), as a new context in which to study generalizations of AdS/CFT and for its role in celestial holography. A basis of time-periodic solutions to the Klein-Gordon wave equation is found in terms of hypergeometric functions. They  fall into a  PT even and a PT odd principal series representation, rather than the more familiar highest-weight representations of the cover of \sot. For masses below the Breitenlohner-Freedman (BF) bound, the modes fall on the unitary principal series.  The presence of CTCs precludes the usual canonical quantization, but geometric quantization, which begins with a symplectic form on the phase space of classical solutions, is applicable. Operators, commutators, an \sot  invariant vacuum and a Fock space are constructed and transform like those of  a CFT$_2$.
The Fock space norm is positive below the BF bound.  In celestial holography, \AdS\ arises as leaves of a hyperbolic foliation of Klein space.  Our  analysis determines  new  entries in the symmetry-constrained celestial bulk-to-boundary dictionary.  In particular the Klein space $\S$-matrix is dual to a maximally entangled state in the tensor product of two copies of the `wedge CFT$_2$' associated to the timelike and spacelike wedges of Klein space. Translation invariance is not present in the wedge CFT$_2$ itself but emerges as a property of this  maximally entangled state.     \end{abstract}

\end{center}

\end{titlepage}

\tableofcontents
\section{Introduction}

The study of quantum field theory (QFT) and quantum gravity  on AdS$_3$ has been a rich source of insight over the decades. In this paper we
consider QFT on the timelike quotient, \AdS, with the period set so that the isometry group is exactly \sot rather than its universal cover. The study of quantum systems on any spacetime with closed timelike curves (CTCs) is  challenging \cite{Deutsch:1991nm,Boulware:1992abc,Hartle:1993sg, Politzer:1994ctc,Bennett:2009rt,Lloyd:2010nt,Akhmedov:2018lkp,Hartle:2021qng,Luminet:2021qae, Bishop:2024cqa}. Here we show that special properties of \AdS\ allow some quantum systems to be defined using  geometric quantization.

\AdS\ arises as hyperbolae of constant radius in   flat signature $(2,2)$ `Klein space' where \sot acts as the Lorentz group \cite{Atanasov:2021oyu}.  4D QFT on Klein space is perturbatively well-defined  and gives  Minkowskian scattering amplitudes  by analytic continuation.  Any 4D theory may be reduced, in the spirit of \cite{deBoer:2003vf},  to a theory on \AdS\ with a tower  of fields  and  a continuum of masses. Doing so gives us examples, albeit with exotic properties, of QFT correlators on \AdS.

This motivates the question of how or when QFT can be defined directly on \AdS\ without reduction from 4D. In Section 3  we consider the simplest -- but already nontrivial -- case of a free complex scalar with mass $m^2$. On AdS$_3$, the   general solutions fall into  highest-weight  representations of \sot with definite asymptotic falloffs, where the total weight is the global  AdS$_3$ energy.  These are not suitably periodic in time.  In \AdS, in contrast we show that the  periodic solutions have mixed falloffs. They fall  into a pair of principal series representations, one PT even and one PT odd.\footnote{The action of PT is defined in \eqref{eq:pt} below.}  We also construct   conformal primary modes which correspond to operator insertions at a point in the Lorentzian torus \t2\ at the boundary of \AdS,   As in AdS$_3$, these obey a highest-weight condition where the weight is dilation weight away from the insertion point,  rather than the global energy.  These are essentially periodic bulk-to-boundary Green functions. A formula is derived  for the  expansion of these Green functions as sums of elements in the principal series representations. 

In AdS$_3$ there is a unitary norm on the space of solutions for $m^2$ above the Breitenlohner-Freedman (BF) bound,  $m^2=-{1\over \ell_{AdS}^2}$ with weight $\D=1+\sqrt{1+m^2\ell_{AdS}^2}$. Below the BF bound  $\Delta=1+i\l$ for real $\l$ and the theory is not unitary. 
Interestingly, in \AdS , the situation is just the opposite. Above the BF bound, the principal series weights 
are real and there is no unitary norm. A unitary norm does exist for masses {\it below} the BF bound where the weights $\Delta=1+i\l$ fall on the unitary principal series.  

The next step is to try to define a quantum theory on \AdS .  The usual canonical quantization procedure begins with  commutation relations between the scalar field and its first time derivative 
on a fixed time slice.  This fails in the presence of  CTCs due to the teleological constraints. Instead in section 4 we adopt geometric quantization \cite{Woodhouse:1980pa}.  This generalization reduces to canonical quantization under suitable conditions. 
It begins with a choice of symplectic form on the classical phase space. We show that there is a unique two-parameter \sot-invariant family of such forms on the phase space of a free complex massive scalar, one each for the PT even and PT odd representations denoted $\alpha_E$ and $\alpha_O$.\footnote{Due to the mixed boundary falloffs, the standard symplectic form constructed from integrating the symplectic  current over a fixed time slice is not conserved and so cannot be used for quantization. This is related to the inapplicability of canonical quantization.} To quantize the theory the modes are promoted to operators with appropriate commutation relations. 
From this we  define an \sot invariant vacuum state and  a Fock space with a norm.  Below the BF bound, for positive $\alpha_{E,O}$,  the norm becomes the unitary one on the unitary principal series.  It is shown that the vacuum two point function of the conformal primary operators matches the two-parameter family of conformally invariant two point functions on \t2.

 An alternate approach to defining a quantum theory is in terms of correlators on the boundary of \AdS, potentially derived from a path integral. This path integral  approach to quantum mechanics in the presence of CTCs is analyzed  in \cite{Hartle:1993sg}. This presumably leads to \AdS -Witten diagrams with 
 time-periodic Green functions of the type  in equation \eqref{btb} below. A choice must  be determined among the  one parameter family of such Green functions characterized by an $i\e$ prescription.\footnote{A priori one may consider linear combinations of $\pm i\e$. This is parallel to the choice of $\alpha_E$ and $\alpha_O$ discussed in 4.1  in the geometric quantization. In  celestial holography applications, a specific choice is singled out and various $N$-point functions have been computed \cite{Melton:2023bjw,Melton:2024gyu}.}
 We expect this ultimately to be the most efficient approach to inclusion of interactions and loop corrections. In this paper however we focus on the  nature of quantum states which is sidestepped in this approach. 
 
 A minimal generalization of the AdS$_3$/CFT$_2$ correspondence suggests that quantum gravity on \AdS\ should be dual to CFT$_2$ on the  boundary Lorentzian torus \t2. 
 In \cite{Melton:2023hiq},  the general fully conformally invariant CFT$_2$ $n$-point correlators with arbitrary non-integral weights on \t2\ were constructed.  If quantum gravity on \AdS\ can be defined the boundary correlators will take these forms. 
 
A primary motivation for this work is the  application to celestial holography.  
Flat Klein space divides into two $\pm$ wedges characterized by the sign, timelike or spacelike,  of  $X^\mu X_\mu$.  Each of these wedges can be foliated by a family of \AdS\ slices characterized by constant radius $X^\mu X_\mu$ \cite{Atanasov:2021oyu}. Field theory (or ultimately quantum gravity) on Klein space may thus be rewritten as a pair of entangled  field theories on two copies of \AdS. The   AdS$_3$/CFT$_2$ dictionary, as minimally adapted to periodic time, identifies the boundary correlators of each of these wedge field theories as a wedge CFT$_2$.\footnote{ These are simply related to the leaf CFTs discussed in \cite{Melton:2023bjw,Melton:2024gyu} by an integral over all the leaves in a wedge. The correlators were referred to as `half amplitudes' in \cite{Melton:2023bjw}. The main difference is a constraint on the net  weight of the operators.}  Together with the analysis of Section 3 and 4, this enables  in Section 5 the construction of a precise bulk-to-boundary dictionary for Klein space involving an entangled pair of wedge CFT$_2$s.\footnote{The dictionary so constructed 
from symmetries alone  gives the operators, states and correlators of the wedge CFT$_2$ in terms of bulk counterparts, but much more is needed for an independent 2D definition of an interacting  wedge CFT$_2$. An example of an independent  leaf dual for MHV scattering amplitudes appears in \cite{Melton:2024gyu}.} In particular using  \cite{Melton:2024pre} we show that the bulk $\S$-matrix is dual to a maximally entangled state of this  pair of wedge CFT$_2$s. 

Wedge CFT$_2$s have 2D conformal invariance which  implies 4D \sot Lorentz invariance. However they are not manifestly translation invariant. In this context translation invariance is explicitly seen to emerge as a property of the maximally entangled state of two copies of the wedge Hilbert space.

\section{Periodic \AdS\   geometry}
This section reviews \AdS\ geometry. The unit radius \AdS\  metric is
\be ds_3^2 =-\cosh^2 \rho dt^2+\sinh ^2 \rho d\theta^2+{d\rho^2},\ee with the identifications 
\be~~~~t\sim t+2\pi,~~ \theta \sim \theta+2\pi.\ee The period of the timelike identification ensures that every bulk null geodesic closes after reflecting twice off the boundary. 
In terms of coordinates
\be \s={t+\theta \over 2},~~~~~\bs={t-\theta \over 2},\ee
 one has 
\be ds_3^2 =-2\cosh 2 \rho d\s d\bs -d\s^2-d\bs^2+{d\rho^2}, ~~~~\s\sim \s+2\pi,~~ (\s,\bs) \sim  (\s+\pi,\bs+\pi).\ee
\AdS\ is depicted in Figure \ref{fig:fig1}.

\begin{figure}[h]
\begin{center}

\tikzset{every picture/.style={line width=0.75pt}} 

\begin{tikzpicture}[x=0.75pt,y=0.75pt,yscale=-1,xscale=1]

\draw    (101,94.5) .. controls (132,105.5) and (123,172.5) .. (100,177.5) ;
\draw   (15,133) .. controls (15,75.84) and (54.18,29.5) .. (102.5,29.5) .. controls (150.82,29.5) and (190,75.84) .. (190,133) .. controls (190,190.16) and (150.82,236.5) .. (102.5,236.5) .. controls (54.18,236.5) and (15,190.16) .. (15,133) -- cycle ;
\draw    (111,93.5) .. controls (76,84.5) and (76,189.5) .. (111,177.5) ;
\draw  [draw opacity=0] (83.72,136.07) .. controls (84.42,136.74) and (84.79,137.45) .. (84.79,138.18) .. controls (84.77,142.87) and (69.36,146.62) .. (50.36,146.57) .. controls (31.36,146.52) and (15.96,142.67) .. (15.98,137.99) .. controls (15.98,137.07) and (16.56,136.2) .. (17.64,135.38) -- (50.38,138.08) -- cycle ; \draw   (83.72,136.07) .. controls (84.42,136.74) and (84.79,137.45) .. (84.79,138.18) .. controls (84.77,142.87) and (69.36,146.62) .. (50.36,146.57) .. controls (31.36,146.52) and (15.96,142.67) .. (15.98,137.99) .. controls (15.98,137.07) and (16.56,136.2) .. (17.64,135.38) ;  
\draw  [draw opacity=0][dash pattern={on 4.5pt off 4.5pt}] (17.06,140.41) .. controls (16.36,139.74) and (15.98,139.04) .. (15.98,138.31) .. controls (15.95,133.62) and (31.32,129.72) .. (50.33,129.6) .. controls (69.33,129.47) and (84.76,133.17) .. (84.79,137.85) .. controls (84.79,138.77) and (84.22,139.65) .. (83.14,140.48) -- (50.38,138.08) -- cycle ; \draw  [dash pattern={on 4.5pt off 4.5pt}] (17.06,140.41) .. controls (16.36,139.74) and (15.98,139.04) .. (15.98,138.31) .. controls (15.95,133.62) and (31.32,129.72) .. (50.33,129.6) .. controls (69.33,129.47) and (84.76,133.17) .. (84.79,137.85) .. controls (84.79,138.77) and (84.22,139.65) .. (83.14,140.48) ;  
\draw  [draw opacity=0] (188.72,138.07) .. controls (189.42,138.74) and (189.79,139.45) .. (189.79,140.18) .. controls (189.77,144.87) and (174.36,148.62) .. (155.36,148.57) .. controls (136.36,148.52) and (120.96,144.67) .. (120.98,139.99) .. controls (120.98,139.07) and (121.56,138.2) .. (122.64,137.38) -- (155.38,140.08) -- cycle ; \draw   (188.72,138.07) .. controls (189.42,138.74) and (189.79,139.45) .. (189.79,140.18) .. controls (189.77,144.87) and (174.36,148.62) .. (155.36,148.57) .. controls (136.36,148.52) and (120.96,144.67) .. (120.98,139.99) .. controls (120.98,139.07) and (121.56,138.2) .. (122.64,137.38) ;  
\draw  [draw opacity=0][dash pattern={on 4.5pt off 4.5pt}] (122.06,142.41) .. controls (121.36,141.74) and (120.98,141.04) .. (120.98,140.31) .. controls (120.95,135.62) and (136.32,131.72) .. (155.33,131.6) .. controls (174.33,131.47) and (189.76,135.17) .. (189.79,139.85) .. controls (189.79,140.77) and (189.22,141.65) .. (188.14,142.48) -- (155.38,140.08) -- cycle ; \draw  [dash pattern={on 4.5pt off 4.5pt}] (122.06,142.41) .. controls (121.36,141.74) and (120.98,141.04) .. (120.98,140.31) .. controls (120.95,135.62) and (136.32,131.72) .. (155.33,131.6) .. controls (174.33,131.47) and (189.76,135.17) .. (189.79,139.85) .. controls (189.79,140.77) and (189.22,141.65) .. (188.14,142.48) ;  
\draw    (168.8,118.8) .. controls (161.56,87.44) and (145.48,67.72) .. (131.51,58.47) ;
\draw [shift={(130,57.5)}, rotate = 31.61] [fill={rgb, 255:red, 0; green, 0; blue, 0 }  ][line width=0.08]  [draw opacity=0] (9.6,-2.4) -- (0,0) -- (9.6,2.4) -- cycle    ;
\draw    (50.38,138.08) -- (83.3,138.76) ;
\draw [shift={(85.3,138.8)}, rotate = 181.18] [fill={rgb, 255:red, 0; green, 0; blue, 0 }  ][line width=0.08]  [draw opacity=0] (9.6,-2.4) -- (0,0) -- (9.6,2.4) -- cycle    ;
\draw    (171,147.5) -- (176.02,146.78) ;
\draw [shift={(178,146.5)}, rotate = 171.87] [fill={rgb, 255:red, 0; green, 0; blue, 0 }  ][line width=0.08]  [draw opacity=0] (9.6,-2.4) -- (0,0) -- (9.6,2.4) -- cycle    ;

\draw (129.1,66.4) node [anchor=north west][inner sep=0.75pt]  [font=\footnotesize]  {$t$};
\draw (66.6,119.4) node [anchor=north west][inner sep=0.75pt]  [font=\footnotesize]  {$\rho $};
\draw (164.1,150.4) node [anchor=north west][inner sep=0.75pt]  [font=\footnotesize]  {$\theta        $};

\end{tikzpicture}

\end{center}
\caption{Geometry of \AdS\label{fig:fig1}}
\end{figure}
The isometries of \AdS\  form the group  \sot, rather than its universal cover as is the case for AdS$_3$. \sot is infinitesimally generated by the Killing vectors
\begin{equation}\label{eq:Lgens}
\begin{split}
&L_{\pm1}  =\frac{e^{\mp 2i\s}}{2}\left(\pm \partial_\rho-i \coth 2\rho \partial_\s-i \operatorname{csch} 2\rho \partial_\bs \right) \\
&L_0  =-\frac{i}{2}\partial_\s \\
&\bar{L}_{\pm1}  =\frac{e^{\mp 2i\bs}}{2}\left(\pm \partial_\rho-i \coth 2\rho \partial_\bs-i \operatorname{csch} 2\rho \partial_\s \right) \\
&\bar{L}_0  =-\frac{i}{2}\partial_\bs.
\end{split}
\end{equation}  
These obey the algebra 
\begin{equation}
    [L_n,L_m] = (n-m)L_{n+m},\ [\bar{L}_n,\bar{L}_m] = (n-m)\bar{L}_{n+m}.
\end{equation}
The center of \sot contains the non-trivial central element $-I$, which we shall refer to as PT. In the coordinates above, its action is \be \label{eq:pt} {\rm PT}:~~~(t,\theta)\to (t+\pi, \theta+\pi),~~~(\s,\bs)\to (\s+\pi, \bs).\ee
The conformal boundary of \AdS\  is a Lorentzian torus (with periodic time) and  metric
\begin{equation}
    ds^2 = -d\s d\bs.
\end{equation}
Null geodesics in the boundary all close after a single period.

\section{Periodic scalar modes} \label{psm}
This section describes  solutions to the Klein-Gordon equation for a complex massive scalar on \AdS\ 
\begin{equation}\label{kge}
    \square \Phi = m^2\Phi,
\end{equation}
where $m^2$ is real but may be positive or negative.
 In an $(L_0, \bar L_0)$ eigenbasis, solutions are given by hypergeometric functions with mixed falloffs at infinity. They  organize   into  one integer and one half-integer principal series representation  of \sot $\sim $ \sllr , in contrast to the highest-weight representations appearing for the cover AdS$_3$.   
We also see that they fall into the unitary  principal series representations for $m^2<-1$.\footnote{Special representations occurring at discrete values of the mass are not considered. } We further decompose local conformal primary solutions, given by bulk-to-boundary propagators, as sums over the principal series representations. 
\subsection{Principal series modes}
We seek a basis of solutions of \eqref{kge} that diagonalize translations $(L_0, \bar L_0)$ around the Lorentzian torus. We accordingly expand the general solution in modes that diagonalize rotations around the compact directions as
\begin{equation} \label{eq:ps}
    \Phi(x) 
    = \sum_{p \pm q \in \mathbb{Z}} c_{p,q}\phi^\D_{p,q}(x),\ \phi^\D_{p,q}(x) = e^{-2ip\s-2iq\bs}\varphi^{\Delta}_{p,q}(\rho),
\end{equation}
where $x = (\rho,\s,\bs)$ are coordinates in the bulk, $(p,q)$ are either both integers or both half integers and $c_{p,q}$ arbitrary constants. The massive wave equation for $\Phi(x)$ implies 
\be [{(p+q)^2 {\rm sech}^2 \rho}-{ (p-q)^2 {\rm csch}^2\rho }+2\coth 2 \rho \p_\rho+ \p_\rho^2]\varphi^{\Delta}_{p,q}(\rho)=\D(\D-2)\varphi^{\Delta}_{p,q}(\rho),
\ee
where we define
\be \D_\pm=1\pm\sqrt{1+m^2},\ee
and, redundantly,  
\be \D=\D  _+=2-\D_-\ee
to avoid notational clutter. 
This is a second order ODE with a unique\footnote{Since $\D_+(\D_+-2)=\D_-(\D_--2)$ one might have expected a second set of solutions with $\D_+\to\D_-$. However as discussed  in Appendix \ref{app:ds} these solutions are proportional to the ones given here and the $\D_\pm$ representations are equivalent.} (up to an overall normalization) solution for every $m^2$
 regular at $\rho=0$. It is  given by the hypergeometric function:
\begin{equation}
    \begin{split}\label{hsl}
    \varphi^{\D}_{p,q}(\rho) &= 2^{{\D}-2}\frac{\Gamma({\D}/2-p)\Gamma({\D}/2+q)}{\Gamma({\D}-1)\Gamma(1-p+q)}\tanh^{q-p}\rho~\sech^{\D}\rho~\pFq{2}{1}{{\D}/2-p,{\D}/2+q}{1-p+q}{\tanh^2\rho}
    \end{split}
\end{equation}
for $q \ge p$; for $q < p$, the regular solution exchanges $p$ and $q$. Near the origin $\rho\rightarrow0$
\be  \varphi^{\D}_{p,q}(\rho) \sim 2^{{\D}-2}\frac{\Gamma({\D}/2-p)\Gamma({\D}/2+q)}{\Gamma({\D}-1)\Gamma(1-p+q)}\rho^{q-p}.\ee
For $q \ge p$, these have no pole at $\rho = 0$. For $q < p$, the smooth solution can be found by exchanging $p$ and $q$. 

Using hypergeometric reflection identities, we obtain the large $\rho$ behavior
\begin{equation}
    \begin{split}
        \varphi^{\D}_{p,q}(\rho) &\sim e^{(\D-2)\rho}+2^{2{\D}-2}\frac{(-1)^{2p}\Gamma(1-{\D})}{\Gamma({\D}-1)}\frac{\Gamma({\D}/2+p)\Gamma({\D}/2+q)}{\Gamma(1-{\D}/2+p)\Gamma(1-{\D}/2+q)}e^{-{\D}\rho}  .
    \end{split}
\end{equation}
The normalization of the modes in \eqref{hsl} was chosen to give this simple asymptotic behavior.\footnote{For $m^2<-1$ the coefficient of the second factor is a pure phase.}  
These solutions necessarily involve both falloffs near the boundary; periodic solutions involving a single falloff at infinity are singular at the origin.

Under the action of the \sot generators defined above, the periodic wavefunctions $\phi^\D_{p,q}(x)$ satisfy
\be\begin{split}\label{tsf}
    L_n\phi^{\Delta}_{p,q}(x) &= [n \frac{\Delta-2}{2}- p]\phi^{\Delta}_{p+n,q}(x)\\
    \bar L_n\phi^{\Delta}_{p,q}(x) &= [n \frac{\Delta-2}{2}- q]\phi^{\Delta}_{p,q+n}(x),
\end{split} \ee
which are the defining relations of principal series representations of \sot. Since integer and half-integer values of $p$ and $q$ do not mix under \eqref{tsf}, we conclude that a basis of solutions of the massive scalar wave equation is provided by the elements of one  PT even and one  PT odd weight $\D=\D_+$ principal series representation of \sot.

For $m^2\ge-1$ so that ${\D_+} \in \mathbb{R}$, $(\phi^{\D_+}_{p,q})^* = \phi^{\D_+}_{-p,-q}$. 
If $m^2 < -1$, so that ${\D_+} \in 1 + i\mathbb{R}$, then $(\phi^{\D_+}_{p,q})^* = \phi^{{\D_-}}_{-p,-q}$.  Such modes are in the unitary principal series.

The structure of these solutions is rather different than what is usually studied in unidentified AdS$_3$. Reference \cite{Maldacena:1998bw} constructs smooth non-periodic solutions with a single falloff at infinity. These form highest-weight rather than principal series representations. Highest and lowest weight representations together  may be recombined into principal series representations, see e.g.\cite{Guijosa:2003ze}.

\subsection{Conformal primary modes}
An alternate  basis of solutions, labeled by a point denoted $(\hs,\hbs)$ on the boundary, rather than global frequencies $(p,q)$, is given by the familiar bulk-to-boundary propagator: 
\bea \label{btb}G^\D(\hat \s,\hat \bs; x )=( \epsilon +i\sin(\s-\hat \s)\sin(\bs-\hat\bs)\re^\rho - i\cos(\s-\hat \s)\cos(\bs-\hat\bs)\re^{-\rho})^{-\Delta}.\eea
 The $i\e$ prescription  given here is manifestly time-periodic. It  differs from that of either the non-periodic time-ordered product or Wightman function
employed on the universal cover, but appeared in \AdS\ correlators in \cite{Melton:2023hiq}. 
 Here and below  we take the branch cut so that
\be \label{bcp}
(\eps-i y)^{-\Delta} = \re^{i\pi\Delta/2}|y|^{-\Delta}\Theta(y) + \re^{-i\pi\Delta/2}|y|^{-\Delta}\Theta(-y)
\ee
for real nonzero $y$.  With this definition $G^\D$ is invariant under CPT defined as  PT ($\s\to\s+\pi$) combined with complex conjugation  with $\D$ held fixed. 

As in the more familiar AdS$_3$ case, the bulk-to-boundary propagator $G^{\D}$ is a conformal primary with respect to the \sot$ \sim~$\sllr\  basis associated to a boundary point $(\hs,\hbs)$ \cite{Atanasov:2021oyu}:
\bea && \label{hf} H_0^{\hs}=\tfrac{1}{2}\big(e^{2i\hs}L_1- e^{-2i\hs}L_{-1}\big),~~~~ H^\hs_{\pm 1}=iL_0\mp\tfrac{i}{2}\big(e^{2i\hs}L_1+ e^{-2i\hs}L_{-1}\big) , 
\cr && \label{hf} \bar H_0^{\hs}=\tfrac{1}{2}\big(e^{2i\hs}\bar L_1- e^{-2i\hs}\bar L_{-1}\big),~~~~ \bar H^\hs_{\pm 1}=i\bar L_0\mp\tfrac{i}{2}\big(e^{2i\hs}\bar L_1+ e^{-2i\hs}\bar L_{-1}\big) . \eea
Explicitly, it  obeys the highest-weight conditions  \bea && \label{gh}H^\hs_{ 1}G^{\D}(\hs,\hbs;x)=0,~~~ H^\hs_{0}G^{\D}(\hs,\hbs;x)=\tfrac{\D}{2}G^{\D}(\hs,\hbs;x),\cr &&\label{gh}\bar H^\hs_{ 1}G^{\D}(\hs,\hbs;x)=0,~~~ \bar H^\hs_{0}G^{\D}(\hs,\hbs;x)=\tfrac{\D}{2}G^{\D}(\hs,\hbs;x).\eea
  Near the \AdS\  boundary  the propagator behaves as 
\be 
\begin{split}
    G^\D(\hat \s,\hat \bs; x )\xrightarrow[]{\rho\rightarrow \infty} &~ (\e+i\sin(\s-\hat \s)\sin(\bs-\hat\bs))^{-\D}\re^{-\D\rho} \\
    & + \frac{2\pi i}{1-\D}[\re^{i\pi\D/2}\d(\s-\hat \s)\d(\bs-\hat \bs)-\re^{-i\pi\D/2}\d(\s-\hat \s+\pi)\d(\bs-\hat \bs)]\re^{(\D-2)\rho}
\end{split} 
\ee
We note that, despite the fact that $G^\D$ is a primary of weight $\D$, there are two points on the boundary with the `shadow falloff' $e^{(\D-2)\rho}$. It follows  that a pair of $G^\D$ at different boundary points can have an associated  flux of the  Klein--Gordon current through the boundary (potentially associated to operator insertions).  This is also the case in AdS$_3$.

Matching the large $\rho$ expansion of the principal series and conformal primary wavefunctions, we see that they are related by 
\be \label{gph} G^\D(\hat \s,\hat \bs; x )= \frac{i}{\pi(1-\Delta)}\sum_{p\pm q \in \mathbb{Z}}[e^{i\pi\D/2}-(-1)^{2p}e^{-i\pi\D/2}]e^{2ip\hs+2iq\hbs}\phi^\D_{p,q}(x).\ee
Hence  a local conformal primary mode decomposes into principal series representations in the global \sot for which $L_0\pm \bar L_0$ have integer eigenvalues.  In the following  $G^\D$ enters  the description of local boundary operators while $\phi^\D_{p,q}$
enters that of  global states. 

\section{Quantization} 

In this section we quantize the classical theory of the preceding section, accounting for the novelties of periodic time.

The most common approach to quantization of a scalar field is the canonical one,
which begins with a Poisson bracket between the phase space coordinates $\Phi (t,\phi,\rho)$ and $\p_t\Phi(t,\phi,\rho)$ on a constant time slice and then evolve states in time with  a Hamiltonian.  
However this approach is problematic  whenever there are CTCs which impose  teleological constraints on the evolution \cite{Deutsch:1991nm,Boulware:1992abc,Hartle:1993sg, Politzer:1994ctc,Bennett:2009rt,Lloyd:2010nt,Akhmedov:2018lkp,Hartle:2021qng,Luminet:2021qae, Bishop:2024cqa}. 
 
 Instead we will use the more general approach of geometric quantization. This  begins with an invertible  symplectic form
$\Omega$ on the phase space of classical solutions of the  complex massive scalar wave equation, which in our case is the complex span of either the principal series or conformal primary  modes described above. In Subsection \ref{csp} (and accompanying Appendix \ref{app:gensp}) we find the most general \sot covariant symplectic form and Poisson bracket on the classical phase space which has two free parameters.  In \ref{oas} we promote the Poisson bracket to a quantum commutator and define an \sot invariant vacuum state and Fock space.  In \ref{sec:2pf} we compute the two point function in this vacuum state. 

In the following  two subsections we restrict to the case $m^2<-1$, so that 
\be \D_\pm=1\pm i\l,~~~\l>0.\ee
This case is of special interest because the quantum states are on the unitary principal series and have a unitary norm as well as the fact that  it arises in celestial holography.  Geometric quantization works similarly for $m^2\ge-1$ and is treated in Subsection \ref{sec:ps}.

\subsection{Classical symplectic product} \label{csp}
 \sot covariance  of the symplectic product requires  that  it obey 
\begin{equation}
    L_n(\phi|\psi) = (L_n\phi|\psi) + (\phi|L_n\psi) = 0
\end{equation}
for $n = -1, 0, 1$. 
A general complex solution  can be expanded in the eigenmodes $\phi_{p,q}^{\D_+}(x) $ defined in \eqref{eq:ps}.
In this basis the \sot covariant  symplectic  product  is  shown  in Appendix \ref{app:gensp} to be 
\begin{equation}\label{sfo}
    (\phi^{\Delta_+}_{p,q}|{\phi}^{\Delta_+}_{r,s}) = -\frac{i}{\alpha_p}\frac{4^{\Delta_+-1}\Gamma(1-\Delta_+)}{\Gamma(\Delta_+-1)}(-1)^{2p}\frac{\Gamma(p+\Delta_+/2)\Gamma(q+\Delta_+/2)}{\Gamma(p+1-\Delta_+/2)\Gamma(q+1-\Delta_+/2)}\delta_{p+r}\delta_{q+s}
\end{equation} 
where 
\begin{equation}
    \alpha_p = \begin{cases}
        \alpha_E & p \in \mathbb{Z} \\
        \alpha_O & p \in \mathbb{Z}+1/2.
        \end{cases}
\end{equation}
These two constants govern the overall magnitude of the PT even and odd symplectic forms. Note that the first argument is a mode of the $\Phi$ field, while the second is a mode of the $\bar{\Phi}$ field. 

This looks simpler if we transform the second argument to the $\D_-$ basis:
\begin{equation}
     (\phi^{\Delta_-}_{p,q}|{\phi}^{\Delta_+}_{r,s}) = \frac{-i}{\alpha_p}\delta_{p+r}\delta_{q+s}.
\end{equation}
\subsection{Operators and states} \label{oas}
The next step is to promote the scalar $\Phi(x)$ and its complex conjugate ${ \bar \Phi}(x)$ to quantum operators 
${\bf \Phi}(x)$ and  ${\bf \bar \Phi}(x)$, distinguished by the boldface type.  We begin by mode-expanding the field operators as \bea\label{dsfg}&&{\bf \Phi}(x)=\sum_{p\pm q \in \mathbb{Z}}{\bf\Phi}^{\Delta_+}_{p,q} \phi^{\Delta_-}_{-p,-q}(x),\cr
&&{\bf \bar \Phi}(x)=\sum_{p\pm q \in\mathbb{Z}}{ \bf\bar\Phi}^{\Delta_-}_{p,q} \phi^{\Delta_+}_{-p,-q}(x).\eea
This mode expansion is chosen so that 
\begin{equation}
    [L_n,\bPhi^{\D_+}_{p,q}] = \left[n\frac{\D_+-2}{2}-p\right]\bPhi^{\D_+}_{p+n,q}.
\end{equation}
Here for future convenience we have expanded ${\bf \bar \Phi}(x)$ in $\phi^{\Delta_+}_{p,q}$ rather than 
$\phi^{\Delta_-}_{p,q}$.
Geometric quantization then posits that the inverse of the classical symplectic product is promoted  to a quantum  operator  commutator according to 
\begin{equation}
    [{\bf\Phi}^{\Delta_+}_{p,q},{\bf{\bar{\Phi}}}^{\Delta_-}_{r,s}] = i(\phi^{\Delta_-}_{-p,-q}|{\phi}^{\Delta_+}_{-r,-s})^{-1} = -\alpha_p\delta_{p+r}\delta_{q+s}.
\end{equation}
We can now define a manifestly \sot-invariant vacuum by 
\begin{equation}
    {\bf{\bar{\Phi}}}^{\Delta_-}_{r,s}\ket{0} = 0\ \ \forall r \pm s \in \mathbb{Z}.
\end{equation}
Defining hermitian conjugation by
\be {\bf\Phi}(x)^\dagger  ={\bf \bar{\Phi}}(x),\ee and 
comparing the mode expansions of ${\bf\Phi}(x)$ and ${\bf \bar{\Phi}}(x)$ we find that 
\begin{equation}\label{eq:dagdef}
    [\bPhi^{\Delta_+}_{p,q}]^\dagger = \bar{\bPhi}^{\Delta_-}_{-p,-q}.
\end{equation}
The conjugate vacuum $\bra{0}$ then obeys  
\begin{equation}
    \bra{0}\bPhi^{\Delta_+}_{p,q} = 0.
\end{equation}
One  finds 
\begin{equation}
    \bra{0}[\bPhi^{\Delta_+}_{p,q}]^\dagger\bPhi^{\Delta_+}_{r,s}\ket{0} = \alpha_p\delta_{p,r}\delta_{q,s}
\end{equation}
yielding  a positive definite inner product for $\alpha_E, \alpha_O > 0$.

In this basis, the Hamiltonian can be written as
\begin{equation}
    H = -\sum_{p\pm q \in\mathbb{Z}}\frac{p+q}{\alpha_p}\bar{\bPhi}^{\D_-}_{p,q}\bPhi^{\D_+}_{-p,-q}.
\end{equation}
Under the conjugation defined by Equation \eqref{eq:dagdef}, $H = H^\dagger$ for $\alpha_{E/O}$ real.
\subsection{The $\Delta \in \mathbb{R}$ case} \label{sec:ps}
Up to now, we have focused on the case $m^2 < -1$, so that $\Delta \in 1+i\mathbb{R}$. If, instead, $m^2 \ge -1$, we would have real $\Delta$. In this case, the principal series modes obey the reality condition
\begin{equation}
    [\phi^{\Delta_\pm}_{p,q}(x)]^* = \phi^{\Delta_\pm}_{-p,-q}(x).
\end{equation}
This implies that conjugation acts on the mode operators by 
\begin{equation}
    [\bPhi^{\Delta_\pm}_{p,q}]^\dagger = \bar{\bPhi}^{\Delta_\pm}_{-p,-q},
\end{equation}
and 
\begin{equation}
      \bra{0}[\bPhi^{\Delta_+}_{p,q}]^\dagger\bPhi^{\Delta_+}_{r,s}\ket{0} = \alpha_p\frac{4^{\Delta_+-1}\Gamma(1-\Delta_+)}{\Gamma(\Delta_+-1)}(-1)^{2p}\frac{\Gamma(p+\Delta_+/2)\Gamma(q+\Delta_+/2)}{\Gamma(p+1-\Delta_+/2)\Gamma(q+1-\Delta_+/2)}\delta_{p,r}\delta_{q,s}.
\end{equation}
There are no choices of $\alpha_p$ such that this is positive definite. 
\subsection{Two-point functions} \label{sec:2pf}
This subsection computes  the two point function in the basis  of  local  operators labeled by $(\hs,\hbs)$:
\begin{equation}\label{eq:pstohweo}
    \begin{split}
       {\bf {\Phi}}^{\Delta_\pm}_E(\hs,\hbs) &= \sum_{p, q \in \mathbb{Z}}  {\bf {\Phi}}^{\Delta_\pm}_{p,q}e^{2ip\hs+2iq\hbs}\\
         {\bf {\Phi}}^{\Delta_\pm}_O(\hs,\hbs) &= \sum_{p, q \in \mathbb{Z}+\frac{1}{2}}  {\bf {\Phi}}^{\Delta_\pm}_{p,q}e^{2ip\hs+2iq\hbs}, \\
    \end{split}
\end{equation}
and similarly for $\bar{\bPhi}$. These create and annihilate PT even or odd modes, which are simple linear combinations of the conformal primary modes, as spelled out in Appendix \ref{append:A}. They  obey highest-weight conditions  of the form \eqref{gh} corresponding to those of   local primary operators in a 2D CFT.
We then have 
\begin{equation}\label{comm}
\begin{split}
  \bra{0}\bar{{\bf\Phi}}^{\Delta_-}_E(\hs_1,\hbs_1){\bf\Phi}^{\Delta_+}_E(\hs_2,\hbs_2)\ket{0} 
  &=[\bar{{\bf\Phi}}^{\Delta_-}_E(\hs_1,\hbs_1),{\bf\Phi}^{\Delta_+}_E(\hs_2,\hbs_2)] \\ 
  &=\pi^2\alpha_E(\delta(\hs_{12})\delta(\hbs_{12}) + \delta(\hs_{12}-\pi)\delta(\hbs_{12})) \\
        \bra{0}\bar{{\bf\Phi}}^{\Delta_-}_O(\hs_1,\hbs_1){\bf\Phi}^{\Delta_+}_O(\hs_2,\hbs_2)\ket{0} 
  &= [\bar{{\bf\Phi}}^{\Delta_-}_O(\hs_1,\hbs_1),{\bf\Phi}^{\Delta_+}_O(\hs_2,\hbs_2) ] \\ 
  &= \pi^2\alpha_O(\delta(\hs_{12})\delta(\hbs_{12}) - \delta(\hs_{12}-\pi)\delta(\hbs_{12})).
 \end{split}
\end{equation}
These are the most general allowed two point function connecting operators of weights $\Delta_+$ and $\Delta_-$ that are PT even and odd respectively.

If we instead consider modes of both $\bPhi$ and $\bar{\bPhi}$ that transform as conformal primaries of weight $\Delta_+$, we have that the PT even part of the two point function is \begin{equation}\label{eq:evtwopoint}
    \begin{split}
        \bra{0}\bar{{\bf\Phi}}^{\Delta_+}_{E}(\hs_1,\hbs_1){\bf\Phi}^{\Delta_+}_E(\hs_2,\hbs_2)\ket{0} 
        &= \frac{\pi4^{\D_+-1}\alpha_E\G(1-\D_+)\G(\D_+/2)^2}{\G(\D_+-1)\G(1/2-\D_+/2)^2}\frac{1}{|\sin\hs_{12}\sin\hbs_{12}|^{\D_+}} \\ 
         &= \frac{\pi(\D_+-1)\alpha_E}{2}\cot\pi\D_+/2\frac{1}{|\sin\hs_{12}\sin\hbs_{12}|^{\D_+}}
    \end{split}
\end{equation}
and the PT odd part is
\begin{equation}\label{eq:oddtwopoint}
     \begin{split}
        \bra{0}\bar{{\bf\Phi}}^{\Delta_+}_{O}(\hs_1,\hbs_1){\bf\Phi}^{\Delta_+}_O(\hs_2,\hbs_2)\ket{0} 
        &=\frac{\pi 4^{\D_+-1}\alpha_O\G(1-\D_+)\G(1/2+\D_+/2)^2}{\G(\D_+-1)\G(1-\D_+/2)^2}\frac{\mathrm{sgn}(\sin\hs_{12}\sin\hbs_{12})}{|\sin\hs_{12}\sin\hbs_{12}|^{\D_+}} \\
    &= \frac{\pi(\D_+-1)\alpha_O}{2}\tan\pi\D_+/2\frac{\mathrm{sgn}(\sin\hs_{12}\sin\hbs_{12})}{|\sin\hs_{12}\sin\hbs_{12}|^{\D_+}}.
    \end{split}
\end{equation}

While $\alpha_E$ and $\alpha_O$ are not fixed by \sot invariance, they may well be fixed by other considerations such as locality in an interacting theory. Indeed we will see below  that in \AdS\  theories arising in  reduction from 4D in celestial holography, which do admit local interactions, 
$\alpha_E=\alpha_O = \mp 1/2\pi\l$. 

\section{Celestial holography}
This section applies the results of the preceding sections to celestial holography as formulated in $(2,2)$-signature Klein space. 
Subsection 5.1 reviews the representation  of a 4D massless  scalar as a continuum of scalars on the \AdS\ leaves of a Klein space foliation.  5.2 uses the description herein of QFT on \AdS\    to give a precise relation between 4D and 2D states and operators and to show that $\alpha_E=\alpha_O=\pm \frac{1}{2\pi i(1-\D)}$ in this context.  In 5.3,  using the results of  \cite{Melton:2024pre},    the 4D $\S$-matrix is identified with  a thermofield double state in the tensor product of 2D wedge CFTs.  

For ease of comparison with the literature in this section we label all fields by
\be \l=i-i\D=i\sqrt{m^2+1},\ee
rather than $\D$.  $\D_+$  ($\D_-$) corresponds to positive (negative) $\l$. 
\subsection{Dimensional reduction Klein space  $\to$   \AdS }

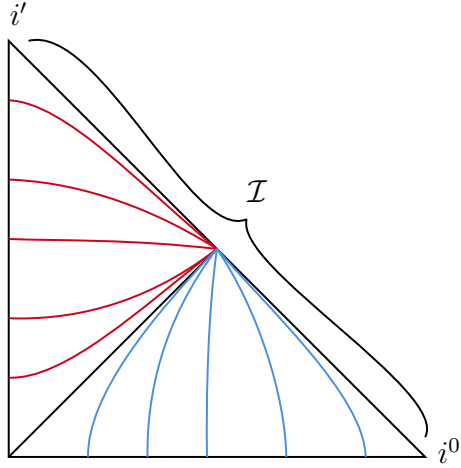
\begin{figure}[h]
\begin{center}

\tikzset{every picture/.style={line width=0.75pt}} 

\begin{tikzpicture}[x=0.75pt,y=0.75pt,yscale=-1,xscale=1]

\draw   (30,30) -- (240,240) -- (30,240) -- cycle ;
\draw    (30,240) -- (135,135) ;
\draw [color={rgb, 255:red, 208; green, 2; blue, 27 }  ,draw opacity=1 ]   (30,170) .. controls (58.5,171) and (96.5,165) .. (135,135) ;
\draw [color={rgb, 255:red, 74; green, 144; blue, 226 }  ,draw opacity=1 ]   (100,240) .. controls (99.5,209) and (107.5,171) .. (135,135) ;
\draw [color={rgb, 255:red, 208; green, 2; blue, 27 }  ,draw opacity=1 ]   (30,200) .. controls (58.5,201) and (96.5,165) .. (135,135) ;
\draw [color={rgb, 255:red, 208; green, 2; blue, 27 }  ,draw opacity=1 ]   (30,130) .. controls (58.5,131) and (90.5,130) .. (135,135) ;
\draw [color={rgb, 255:red, 208; green, 2; blue, 27 }  ,draw opacity=1 ]   (30,100) .. controls (58.5,101) and (95.5,110) .. (135,135) ;
\draw [color={rgb, 255:red, 208; green, 2; blue, 27 }  ,draw opacity=1 ]   (30,60) .. controls (58.5,61) and (97.5,104) .. (135,135) ;
\draw [color={rgb, 255:red, 74; green, 144; blue, 226 }  ,draw opacity=1 ]   (70,240) .. controls (69.5,209) and (102.5,169) .. (135,135) ;
\draw [color={rgb, 255:red, 74; green, 144; blue, 226 }  ,draw opacity=1 ]   (130,240) .. controls (129.5,209) and (130.5,167) .. (135,135) ;
\draw [color={rgb, 255:red, 74; green, 144; blue, 226 }  ,draw opacity=1 ]   (170,240) .. controls (169.5,209) and (155.5,165) .. (135,135) ;
\draw [color={rgb, 255:red, 74; green, 144; blue, 226 }  ,draw opacity=1 ]   (210,240) .. controls (209.5,209) and (159.5,164) .. (135,135) ;
\draw    (40,30) .. controls (75.5,21) and (122.5,132) .. (150,120) ;
\draw    (150,120) .. controls (143.5,144) and (252.5,207) .. (240,230) ;

\draw (245,229) node [anchor=north west][inner sep=0.75pt]    {$i^{0}$};
\draw (29,9) node [anchor=north west][inner sep=0.75pt]    {$i'$};
\draw (149,99) node [anchor=north west][inner sep=0.75pt]    {$\mathcal{I}$};

\end{tikzpicture}

\end{center}
\caption{A toric Penrose diagram of Klein space. The red lines are \AdS\ leaves of the foliation of the $-$ wedge with $X^2 < 0$, while the blue lines are those of  the $+$ wedge with $X^2 > 0$. \label{fig:kleinspace}}
\end{figure}

Celestial holography can be naturally formulated in  Klein space, with metric
\begin{equation}
    \begin{split}
        ds^2 &= -dX_0^2 - dX_1^2 + dX_2^2 + dX_3^2. 
        \end{split}
        \end{equation}
        In coordinates where $X = \tau \hat{x}_\pm(\rho,\s,\bs)$ with $\hat{x}_\pm^2 = \mp 1$ \cite{Melton:2024pre},
        \begin{equation}
            \begin{split}
        ds^2 &= \begin{cases}
                -d\tau^2 + \tau^2ds_3^2 & X^2 < 0 \\
                d\tau^2-\tau^2ds_3^2 & X^2 > 0
        \end{cases}
\end{split}
\end{equation}
where $ds_3^2$ is the standard metric on \AdS. The geometry of Klein space is depicted in Figure \ref{fig:kleinspace}.

By reducing the theory along the $\tau$ direction in the timelike (+) and spacelike ($-$) wedges of Klein space, we can relate theories in 4D flat spacetimes to an interacting pair of theories on \AdS\ with a tower of reduced fields  \cite{deBoer:2003vf,Melton:2024pre}. For a 4D massless scalar, the reduced theory in each region possesses a continuum of fields $\Phi_\lambda$, $\lambda \in \mathbb{R}$, with masses $m^2 = -1 - \lambda^2$. In terms of the field operator, this takes the form 
\begin{equation}\label{eq:ksme}
    \bPhi(\tau\hat{x}_\pm) = \int_{-\infty}^\infty \frac{d\lambda}{2\pi}\tau^{-1-i\lambda}\bPhi^\lambda_\pm(\rho,\s,\bs) = \int_{-\infty}^\infty \frac{d\lambda}{2\pi}\tau^{-1-i\lambda} \sum_{p\pm q\in \mathbb{Z}}\bPhi^{-\lambda}_{p,q,\pm}\phi^{1+i\lambda}_{-p,-q}(\rho,\s,\bs).
\end{equation}
These modes can be combined into the PT even and odd conformal primary operators\footnote{Note that the operator normalization here differs from that of  \cite{Melton:2024pre}. For the exact relationship between the operators defined here and those in \cite{Melton:2024pre}, see  Appendix \ref{app:transconv}.}
\begin{equation}\label{bopb}
    \begin{split}
        {\bf\Phi}^\lambda_{E\pm }(\hs,\hbs)&=  \sum_{p, q \in \mathbb{Z}} {\bf {\Phi}}^\lambda_{p,q,\pm}e^{2ip\hs+2iq\hbs} ,\\
  {\bf\Phi}^\lambda_{O\pm }(\hs,\hbs)&=\sum_{p, q \in \mathbb{Z}+1/2} {\bf {\Phi}}^\lambda_{p,q,\pm}e^{2ip\hs+2iq\hbs}.
\end{split}
\end{equation}
The standard 4D canonical commutators of these operators were computed in  \cite{Melton:2024pre} and found to be\begin{equation}\label{commb}
\begin{split}
[{{\bf\Phi}}^{\l_1}_{E \pm}(\hs_1,\hbs_1),{\bf\Phi}^{\l_2}_{E\pm}(\hs_2,\hbs_2)]  &= \mp\frac{\pi}{2\lambda_1}\delta(\lambda_1+\lambda_2)(\delta(\hs_{12})\delta(\hbs_{12}) + \delta(\hs_{12}-\pi)\delta(\hbs_{12})\\
 [{{\bf\Phi}}^{\l_1}_{O\pm}(\hs_1,\hbs_1),{\bf\Phi}^{\l_2}_{O\pm}(\hs_2,\hbs_2) ] &=\mp\frac{\pi}{2\l_1}\delta(\l_1+\l_2)(\delta(\hs_{12})\delta(\hbs_{12})-\delta(\hs_{12}-\pi)\delta(\hbs_{12})).
 \end{split}
\end{equation}
Translating this into the principal series mode commutators yields
\begin{equation}
    \begin{split}
[\bPhi^{\lambda_1}_{p,q,\pm},\bPhi^{\lambda_2}_{r,s,\pm}] &= \mp\frac{1}{2\pi\lambda_1}\delta(\lambda_1+\lambda_2)\delta_{p+r}\delta_{q+s}.
    \end{split}
\end{equation}
\sot invariant bulk vacua $|0_\pm\rangle$ for the $\pm$ wedges are  then defined by 
\bea && {\bf\Phi}^{\l}_{E+}(\hs,\hbs) | 0_+ \rangle=0,~~~~ {\bf\Phi}^{\l}_{O+}(\hs,\hbs) | 0_+ \rangle=0;  \cr && {\bf\Phi}^{-\l}_{E-}(\hs,\hbs) | 0_- \rangle=0,~~~~ {\bf\Phi}^{-\l}_{O-}(\hs,\hbs) | 0_- \rangle=0 ~~~~\forall \l>0. \eea
A Fock space  is constructed by the action of the other operators  on $|0_-\rangle|0_+\rangle$. 

\subsection{\AdS\ $\to$ wedge CFT$_2$}
The analyses of the  previous sections now enable a precise identification of these 4D bulk Klein space operators and states with two copies (one for each wedge) of operators and states in the principal series on the  \AdS. These  in turn, in a minimal extension of  the usual AdS/CFT bulk-boundary dictionary,  are identified with operators and states in the  wedge CFT$_2$ on \t2\ dual to the bulk \AdS\ theory. This entry in the celestial bulk-to-boundary dictionary is completely determined bottom-up by the symmetries. In anticipation we have already chosen our notation accordingly: \textit{e.g.} ${\bf\Phi}^{\l}_{E+}(\hs,\hbs)$ stands for either a  4D bulk operator 
or its 2D boundary counterpart.

Demanding complete agreement, including between 2D and 4D correlators and commutators,   fixes the arbitrary constants $\alpha_E$ and $\alpha_O$ appearing in the symplectic form \eqref{sfo}. Comparing 
\eqref{comm} and \eqref{commb} one finds, treating $\bPhi^{-\lambda} = \bar{\bPhi}^\lambda$, $\lambda > 0$
\be \alpha_{E\pm} = \alpha_{O\pm} = \mp\frac{1}{2\pi\l}  \ee
while the $\d(\l_1+\l_2)$ factors account for the fact that the 4D theory leads to a continuum rather than a discrete set of scalar fields with varying mass on \AdS.

\subsection{Emergence of spacetime translations}
Further entries in the celestial bulk-to-boundary dictionary are implied by the bulk construction of the Poincaré-invariant  no-boundary state defined in\cite{Melton:2024pre}. Because future and past null boundaries of Minkowski space analytically continue to the single null infinity of Klein space, this state is actually the continuation of the Minkowski $\S$-matrix (viewed as a state in the tensor product of the in and out Hilbert spaces) rather than the Minkowski vacuum.  Since the boundary of Klein space spans the asymptotic boundaries of both wedges, it is a state in the tensor product of the $\pm$ wedge Hilbert spaces.\footnote{The tensor product of the  $\pm$ Hilbert spaces is identified with the tensor product of the  in and out Minkowski Hilbert spaces \cite{Crawley:2021ivb}.} The explicit expression is 
\begin{equation}\label{entx}
    \begin{split}
        \ket{0_K} &= {\cN}e^X\ket{0_+}\ket{0_-} \\
        X &= -\frac{1}{\pi}\int_0^\infty \lambda d\lambda \int_{T^{1,1}}d^2\hs[{\bf\Phi}^{\lambda}_{E-}{\bf\Phi}^{-\lambda}_{E+}+{\bf\Phi}^{\lambda}_{O-}{\bf\Phi}^{-\lambda}_{O+}] \\
        &= -2\pi\int_0^\infty \lambda d\lambda \sum_{p\pm q\in\mathbb{Z}} \bPhi^{\lambda}_{p,q,-}\bPhi^{-\lambda}_{-p,-q,+}
    \end{split}
\end{equation}
where $\cN$ is a normalization constant.\footnote{Reminiscent  relations were found in dS/CFT in \cite{Cotler:2023xku} and in Minkowski  space in \cite{cmsw}. This also resembles the construction of the Minkowski vacuum as a Rindler thermofield double. }
This is a maximally entangled state or, equivalently, a thermofield double at $T\to\infty$. Tracing over the - Hilbert space gives the + identity operator
\be \tr_- \ket{0_K}\bra{0_K}={\bf{1}}_+\ee
for a suitable  choice of $\cN$ \cite{Melton:2024pre}.

This has a nice interpretation. Celestial holography posits that  the flat space $\S$-matrix can be usefully rewritten  as correlators in 
an (exotic and non-unitary) celestial CFT$_2$ (CCFT). The primary motivation is the subleading soft graviton theorem which implies that the quantum gravity $\S$-matrix transforms covariantly under the full local conformal group acting on the celestial sphere. This would seemingly bring the power of CFT$_2$ to bear on 4D quantum gravity.  However,  exotic properties of CCFTs  render many of the standard tools of CFT$_2$ inapplicable. Prominent among these exotic properties is the translational  symmetry of CCFT, which relates  operators whose dimensions differ by an integer and forces the low point correlators to be distributional \cite{Pasterski:2017ylz,Law:2019glh}.  

In the 4D-2D relations considered here, the primary object is the wedge CFT which lives at the boundary of \AdS .  As shown  in \cite{Melton:2023bjw}, the  distributional  low-point celestial correlators are built out of certain linear combinations of smoother wedge correlators. This is possible because 
translations do not preserve the less-exotic  wedge CFT correlators which in some cases have relatively non-exotic (although still non-unitary) realizations  \cite{Melton:2024gyu}.  The wedge  vacuum is conformally but not Poincaré invariant. Translation invariance emerges only at the level of the thermofield double  \eqref{entx} from quantum entanglement.  
\section*{Acknowledgements}

The authors would like to thank Chris Akers, Adam Ball, Tom Hartman, Ilia Kochergin, Ana-Maria Raclariu, Atul Sharma, Diandian Wang and Zixia Wei for useful conversations. This work was supported in part by NSF PHY-2207659, the Simons Collaboration on Celestial Holography, and the Harvard Society of Fellows.
\appendix
\section{Translating Between Principal Series and Conformal Primary Operators} \label{append:A}
Because both the conformal primary modes and the principal series modes provide complete bases for time-periodic solutions to the Klein-Gordon equation on \AdS, the quantum field operator can be expanded in both bases:
\begin{equation}
    \bPhi(x) = \sum_{p \pm q \in \mathbb{Z}}\bPhi^{\Delta_\pm}_{p,q}\phi^{\Delta_\mp}_{-p,-q}(x) = \int d^2\hs \bPhi^{\Delta_\pm}(\hs,\hbs)G^{\D_\mp}(\hat \s,\hat \bs; x ).
\end{equation}
where $\bPhi^{\Delta}_{pq}$ and $\bPhi^{\Delta}(\hs,\hbs)$ are operators. Using the relationship between the principal series and conformal primary modes given by \eqref{gph}, the associated operators are related by 
\begin{equation}
    \begin{split}
        \bPhi^{\Delta_\pm}(\hs,\hbs) &= -\frac{\Delta_\pm-1}{4\pi\sin\pi\Delta_\pm}\sum_{p \pm q \in \mathbb{Z}}(e^{-i\pi\Delta_\mp/2}+(-1)^{2p}e^{i\pi\Delta_\mp/2})e^{2ip\hs+2iq\hbs}\bPhi^{\Delta_\pm}_{p,q} \\
        &= \frac{1-\D_\pm}{4\pi}(\csc\pi\D_\pm/2\bPhi^{\D_\pm}_E - i\sec\pi\D_\pm/2\bPhi^{\D_\pm}_O) 
    \end{split}
\end{equation}
where $\bPhi_E$ and $\bPhi_O$ are defined in  \eqref{eq:pstohweo}. Inverting this, we have that 
\begin{equation}
    \begin{split}
        \bPhi^{\Delta_\pm}_E(\hs,\hbs) &= -\frac{2\pi\sin\pi\Delta_\pm/2}{\Delta_\pm-1} (\bPhi^{\Delta_\pm}(\hs,\hbs) + \bPhi^{\Delta_\pm}(\hs+\pi,\hbs)) \\
        \bPhi^{\Delta_\pm}_O(\hs,\hbs) &= \frac{2\pi\cos\pi\Delta_\pm/2}{i(\Delta_\pm-1)}(\bPhi^{\Delta_\pm}(\hs,\hbs) - \bPhi^{\Delta_\pm}(\hs+\pi,\hbs)).
    \end{split}
\end{equation}
\section{General Symplectic Product}\label{app:gensp}
In this appendix, we find the most general \sot-invariant symplectic product on the phase space of free massive scalars on \AdS. \sot-invariance implies 
\begin{equation}
    L_n(\phi|\psi) = (L_n\phi|\psi) + (\phi|L_n\psi) = 0
\end{equation}
for $n = -1, 0, 1$. Mode expanding our fields in principal series modes, let
\begin{equation}
    (\phi^{\Delta_+}_{A,p,q}|\phi^{\Delta_+}_{B,r,s}) = c^{AB}_{pq,rs}
\end{equation}
where $A$ may label a species of particle. Then, 
\begin{equation}
    L_0(\phi^{\Delta_+}_{A,p,q}|\phi^{\Delta_+}_{B,r,s}) = -(p+r)c^{AB}_{pq,rs} = 0
\end{equation}
so that $c^{AB}_{pq,rs} = c^{AB}_{pq}\delta_{p+r}\delta_{q+s}$. Then,
\begin{equation}
    L_1(\phi^{\Delta_+}_{p,q}|\phi^{\Delta_+}_{r,s}) = \left[\frac{\Delta_+-2}{2}-p\right]c^{AB}_{p+1q}\delta_{p+r+1}\delta_{q+s} + \left[\frac{\Delta_+-2}{2}-r\right]c^{AB}_{pq}\delta_{p+r+1}\delta_{q+s} = 0.
\end{equation}
This fixes
\begin{equation}
    c^{AB}_{pq} = -\frac{i}{\alpha_p^{AB}}\frac{(-1)^{2p}2^{2\D_+-2}\G(1-\D_+)\Gamma(p+\Delta_+/2)\Gamma(q+\Delta_+/2)}{\G(\D_+-1)\Gamma(p+1-\Delta_+/2)\Gamma(q+1-\Delta_+/2)}.
\end{equation}
where 
\begin{equation}
    \alpha_p^{AB} = \begin{cases}
        \alpha^{AB}_E & p \in \mathbb{Z} \\
        \alpha^{AB}_O & p \in \mathbb{Z}+1/2.
    \end{cases}
\end{equation} 
For this to be appropriately antisymmetric, we must have $\alpha^{AB}_p = -\alpha^{BA}_p$ because the ratio of gamma functions is invariant under $(p,q) \to (-p,-q)$. This implies that we need at least two species of scalars to have an \sot invariant symplectic product. Recognizing the ratio of Gamma functions as implementing the discrete shadow discussed in appendix \ref{app:ds}, this implies the symplectic product between principal series and discrete shadowed principal series modes
\begin{equation}
    (\phi^{\Delta_+}_{Apq}|\phi^{\Delta_-}_{Brs}) =-\frac{i}{\alpha_p^{AB}}\delta_{p+r}\delta_{q+s}
\end{equation}
\section{Two-Point Function Sums}
In this section we evaluate the sum in \eqref{eq:evtwopoint} and \eqref{eq:oddtwopoint}:
\begin{equation}
    \begin{split}
        I_+ &= \sum_{p, q \in \mathbb{Z}}  \frac{\G(\Delta/2+p)\G(\Delta/2+q))}{\G(1-\Delta/2+p)\G(1-\Delta/2+q)}e^{2ip\hs_{12}+2iq\hbs_{12}} = S_+(\hs_{12})S_+(\hbs_{12}) \\
        I_- &=\sum_{p,q \in \mathbb{Z}+1/2} \frac{\G(\Delta/2+p)\G(\Delta/2+q))}{\G(1-\Delta/2+p)\G(1-\Delta/2+q)}e^{2ip\hs_{12}+2iq\hbs_{12}} = S_-(\hs_{12})S_-(\hbs_{12})
        \end{split}
\end{equation}
where 
\begin{equation}
\begin{split}
    S_+(\s) &= \sum_{p \in \mathbb{Z}}\frac{\G(\D/2+p)}{\G(1-\D/2+p)}e^{2ip\s} \\
     S_-(\s) &= \sum_{p \in \mathbb{Z}+1/2}\frac{\G(\D/2+p)}{\G(1-\D/2+p)}e^{2ip\s}
    \end{split}
\end{equation}
By symmetry, we know that 
\begin{equation}
    S_+(\s) = \alpha_+ |\sin\s|^{-\Delta},\ S_-(\s) = \alpha_- \mathrm{sgn}(\sin\s)|\sin\s|^{-\Delta}.
\end{equation}
To fix the coefficient, note that 
\begin{equation}
    \begin{split}
        \int_0^{2\pi}S_+(\s)d\s &= \alpha_+\int_0^{2\pi}|\sin\s|^{-\Delta} \\
        2\pi\frac{\G(\D/2)}{\G(1-\D/2)} &= \alpha_+\frac{2\sqrt{\pi}\G(1/2-\D/2)}{\G(1-\D/2)} \\
        S_+(\sigma) &= \frac{\sqrt{\pi}\Gamma(\D/2)}{\G((1-\D)/2)}\frac{1}{|\sin\s|^\Delta}.
    \end{split}
\end{equation}
And 
\begin{equation}
\begin{split}
    \int_0^{2\pi} e^{-i\s}S_-(\s)d\s &= \alpha_-\int_0^{2\pi}e^{-i\s}\mathrm{sgn}(\sin\s)|\sin\s|^{-\Delta} \\
    -2\pi\frac{\Gamma(\D/2+1/2)}{\G(3/2-\D/2)} &= \alpha_-\frac{2i\sqrt{\pi}\G(1-\D/2)}{\G(3/2-\D/2)} \\
    S_-(\sigma) &= \frac{i\sqrt{\pi}\G(1/2+\D/2)}{\G(1-\D/2)}\frac{\mathrm{sgn}(\sin\s)}{|\sin\s|^{\Delta}}.
\end{split}
\end{equation}
Altogether, we have that 
\begin{equation}
    \begin{split} 
    I_+ &= \frac{\pi\G(\D/2)^2}{\G(1/2-\D/2)^2}\frac{1}{|\sin\hs_{12}\sin\hbs_{12}|^\Delta} \\
    I_- &= -\frac{\pi\G(1/2+\D/2)^2}{\G(1-\D/2)^2}\frac{\mathrm{sgn}(\sin\hs_{12}\sin\hbs_{12})}{|\sin\hs_{12}\sin\hbs_{12}|^\Delta}
\end{split}
\end{equation}
\section{Discrete Shadow}\label{app:ds}
In this section we show that principal series representations of (holomorphic) weight $h$ and $1-h$ are equivalent. 

A weight $h$ state in the principal series obeys
\be L_n\Phi^{h}_{p} = [n (h-1)- p]\Phi^{h}_{p+n} \ee
where $h\in{\mathbb{C}}$ and $p\in {\mathbb{Z}}$. Now let us define 
\be \tilde \Phi^{1-h}_p= \frac{\G(1-h+p)}{\G(h+p)}\Phi^h_p.\ee
This transforms as 
\be L_n\tilde \Phi^{1-h}_{p} = [-nh- p]\tilde\Phi^{1-h}_{p+n} \ee
and hence has weight $1-h$ as indicated by the notation. Extending this to our wave functions, we have that 
\begin{equation}
    \phi^{\Delta_-}_{p,q} = 2^{2-2\Delta}\frac{(-1)^{2p}\G(\Delta-1)\G(1+p-\Delta/2)\G(1+q-\Delta/2)}{\G(1-\Delta)\G(p+\Delta/2)\G(q+\Delta/2)}\phi^{\Delta_+}_{p,q}
\end{equation}
transforms as a principal series mode with weight $\Delta_-$.

\section{Evaluating the Clock Product}\label{app:transconv}
In \cite{Melton:2024pre}, operators were defined by taking the clock product between a solution of the wave equation and the bulk field operator:\footnote{We have chosen the notation here to avoid confusion with the operators $\bPhi^\D(\hs)$ defined above.}
\begin{equation}
    \O^\l_{E/O,\pm}(\hs,\hbs) = (\Phi^\l_{E/O}(x|\hs,\hbs)|\bPhi(X))_{\pm}
\end{equation}
where $\Phi^\D_E$ ($\Phi^\D_O$) are the parity even (odd) conformal primary wavefunctions, with $X = \tau \hat{x}_\pm(\rho,\s,\bs)$ in the timelike and spacelike wedges:
\begin{equation}
    \begin{split}
        \Phi^\l_E(X) &= \frac{\sqrt{\tanh\pi|\l|/2}}{8\pi^2}\frac{\G(1+i\l)}{\tau^{1+i\l}}(G^{1+i\l}(\hs,\hbs;x) + G^{1+i\l}(\hs+\pi,\hbs;x)) \\
         \Phi^\l_O(X) &= \frac{\sqrt{\coth\pi|\l|/2}}{8\pi^2}\frac{\G(1+i\l)}{\tau^{1+i\l}}(G^{1+i\l}(\hs,\hbs;x) - G^{1+i\l}(\hs+\pi,\hbs;x)).
        \end{split}
\end{equation}
where $G^\D$ is the conformal primary wavefunction defined in Equation \eqref{btb}. Inserting our mode expansion in Equations \eqref{eq:ksme} and \eqref{bopb} and using the form of the clock product in the timelike and spacelike wedges in \cite{Melton:2024pre}, we see that 
\begin{equation}
    \O^\l_{E/O,\pm}(\hs) = A^\l_{E/O\pm} \bPhi^{\l}_{E/O\pm}(\hs)
\end{equation}
where 
\begin{equation}
    \begin{split}
    A^\l_{E\pm} &= \pm \sgn(\l)\frac{\l\sech\pi\l/2\sqrt{\coth\pi|\l|/2}}{\G(1-i\l)} \\
    A^\l_{O\pm} &= \mp\sgn(\l)\frac{\l\csch\pi\l/2\sqrt{\tanh\pi|\l|/2}}{\G(1-i\l)}.
    \end{split}
\end{equation}
Inserting these relations into the clock product 
\begin{equation}
\begin{split}
    [\O^{\l_1}_{E\pm},\O^{\l_2}_{E\pm}]_\pm &= \mp\sgn(\l_1)\delta(\l_1+\l_2)(\delta(\hs_{12})\delta(\hbs_{12}) + \delta(\hs_{12}-\pi)\delta(\hbs_{12})) \\
    [\O^{\l_1}_{O\pm},\O^{\l_2}_{O\pm}]_\pm &= \pm\sgn(\l_1)\delta(\l_1+\l_2)(\delta(\hs_{12})\delta(\hbs_{12}) - \delta(\hs_{12}-\pi)\delta(\hbs_{12}))
\end{split}
\end{equation}
implies that $\bPhi^\l_{E/O\pm}(\hat{s})$ obeys
\begin{equation}
    \begin{split}
        [\bPhi^{\lambda_1}_{E\pm},\bPhi^{\lambda_2}_{E\pm}]_\pm &= \mp\frac{1}{A_{E\pm}^{\lambda_1}A_{E\pm}^{\lambda_2}}\sgn(\lambda_1)\delta(\lambda_1+\lambda_2)(\delta(\hs_{12})\delta(\hbs_{12}) + \delta(\hs_{12}-\pi)\delta(\hbs_{12})) \\
        &= \mp \frac{\pi}{2\lambda_1}\delta(\lambda_1+\lambda_2)(\delta(\hs_{12})\delta(\hbs_{12}) + \delta(\hs_{12}-\pi)\delta(\hbs_{12})) \\
         [\bPhi^{\lambda_1}_{O\pm},\bPhi^{\lambda_2}_{O\pm}]_\pm &= \pm\frac{1}{A_{O\pm}^{\lambda_1}A_{O\pm}^{\lambda_2}}\sgn(\lambda_1)\delta(\lambda_1+\lambda_2)(\delta(\hs_{12})\delta(\hbs_{12}) - \delta(\hs_{12}-\pi)\delta(\hbs_{12}))\\
         &= \mp\frac{\pi}{2\lambda_1}\delta(\lambda_1+\lambda_2)(\delta(\hs_{12})\delta(\hbs_{12}) - \delta(\hs_{12}-\pi)\delta(\hbs_{12}))
    \end{split}
\end{equation}
and the no-boundary state takes the form
\begin{equation}
    \begin{split}
        \ket{0_K} &= \mathcal{N}e^X\ket{0_+}\ket{0_-}\\
        X &= \frac{1}{2}\int_0^\infty d\lambda\int d^2\hs(\O^\lambda_{-,E}\O^{-\lambda}_{+,E} - \O^{\lambda}_{-,O}\O^{-\lambda}_{+,O} ) \\
        &= -\frac{1}{\pi}\int_0^\infty \lambda d\lambda \int d^2\hs(\bPhi^{\lambda}_{-E}\bPhi^{-\lambda}_{+E} + \bPhi^{\lambda}_{-O}\bPhi^{-\lambda}_{+O}) \\
        &= -2\pi\int_0^\infty \lambda d\lambda \sum_{p\pm q\in\mathbb{Z}} \bPhi^{\lambda}_{p,q,-}\bPhi^{-\lambda}_{-p,-q,+}
        \end{split}
\end{equation}

\bibliographystyle{JHEP}
\bibliography{refs}


\end{document}